\documentclass[aps, preprint, groupedaddress, floatfix, longbibliography]{revtex4-2}

\usepackage{epsfig}
\usepackage{epstopdf}
\usepackage{graphicx}
\usepackage{amsmath}
\usepackage{color}
\usepackage[export]{adjustbox}
\usepackage[colorlinks=true, allcolors=blue]{hyperref}
\usepackage{soul}

\begin{document}

\title{Nearly perfect Fermi surface nesting in hole-doped La$_3$Ni$_2$O$_7$ enables bulk superconductivity without pressure or strain}

\author{Chengliang Xia$^1$, Jiale Chen$^{1,2}$, Hongquan Liu$^{3}$ and Hanghui Chen$^{1,4}$\footnote{Correspondence to: hanghui.chen@nyu.edu}}
\affiliation{$^1$NYU-ECNU Institute of Physics, NYU Shanghai, Shanghai 200124, China\\
  $^2$Key Laboratory of Polar Materials and Devices, Ministry of Education, East China Normal University, Shanghai 200241, China\\
  $^3$Department of Physics, Brown University, Providence, RI 02912, USA\\
  $^4$Department of Physics, New York University, New York, New York 10012, USA}

\date{\today}

\begin{abstract}
  The discovery of high-temperature superconductivity in Ruddlesden-Popper nickelates has drawn great attention. However, unlike cuprates and iron-based superconductors, Ruddlesden-Popper nickelates exhibit superconductivity either under high pressure in bulk samples or under compressive strain in thin films. Genuine bulk superconductivity under ambient pressure has remained elusive in these materials, precluding key measurements such as specific heat and superfluid density. In this work, we combine density-functional-theory, dynamical-mean-field-theory, and random-phase-approximation to solve the superconducting gap equation for bulk hole-doped bilayer nickelate La$_{3-x}$Sr$_x$Ni$_2$O$_7$ at ambient pressure. We find that hole doping induces a Ni-$d_{3z^2-r^2}$-derived $\gamma$ pocket on the Fermi surface, and serves as a tuning parameter for both its size and \textit{shape}. As $x$ approaches 0.4, the $\gamma$ pocket evolves from circular to diamond-shaped and expands to span half of the Brillouin zone, resulting in nearly perfect Fermi surface nesting with the optimal nesting vector $\textbf{Q} = (\pi, \pi)$. This, in turn, strongly enhances antiferromagnetic spin fluctuations and substantially increases the leading superconducting eigenvalue to a level at which superconductivity becomes experimentally observable. Our work provides both a robust mechanism and an experimentally feasible route to inducing the long-sought bulk superconductivity in La$_3$Ni$_2$O$_7$ without pressure or strain.
\end{abstract}

\maketitle


\newpage

\section*{Introduction}

Ruddlesden-Popper (RP) nickelates have recently joined the family of unconventional superconductors~\cite{327nature,zhang2024high,10.1093/nsr/nwaf220,wang2024bulk,cpl_41_1_017401,Zhu2024,PhysRevX.15.021005,ko2025signatures,zhou2025ambient,hao2025superconductivity,qrkk-l2ng,liu2025superconductivity,yang2024orbital}. However, despite their high superconducting transition temperature (the most recent record approaches 100 K~\cite{li2026bulk}), these materials exhibit superconductivity only under high pressure in bulk samples~\cite{327nature,zhang2024high,wang2024bulk,10.1093/nsr/nwaf220,cpl_41_1_017401,Zhu2024,PhysRevX.15.021005} ($> 10$ GPa) or under compressive strain in thin films~\cite{ko2025signatures,zhou2025ambient,hao2025superconductivity,qrkk-l2ng,liu2025superconductivity}. This distinguishes them from cuprates and iron-based superconductors, which exhibit bulk superconductivity under ambient pressure~\cite{bednorz1986possible,PhysRevLett.58.908,schilling1993superconductivity,doi:10.1021/ja800073m,PhysRevLett.101.107006}. The pressure or strain imposed on RP nickelates precludes many key bulk probes of superconductivity, such as specific heat, thermal conductivity, and superfluid density~\cite{eremets2022high,10.1063/5.0061628,Wang_2024}. The absence of bulk superconductivity under ambient pressure in RP nickelates remains an open question~\cite{doi:10.7566/JPSJ.93.053702,liu2023evidence,XIE20243221,chen2024electronic,PhysRevLett.132.256503,khasanov2025pressure,ZHAO20251239,PhysRevB.110.205122,shi2025spin,ni2025spin,PhysRevB.111.184401,Shi_2025}. Under ambient pressure, RP nickelates, such as La$_3$Ni$_2$O$_7$ and La$_4$Ni$_3$O$_{10}$, crystallize in an orthorhombic structure with strong oxygen octahedral rotations, which are believed to suppress the interlayer coupling and thus superconductivity~\cite{PhysRevLett.132.146002,PhysRevB.108.L140504,Maier2026}. However, a recent experiment synthesized a high-symmetry tetragonal structure of nickelates and still did not observe superconductivity, which is ascribed to the absence of spin fluctuations~\cite{shi2025spin}. Whether structural or magnetic factor plays a more decisive role in the absence of bulk superconductivity under ambient pressure in RP nickelates has been intensively debated. In addition to elucidating this issue, it is highly desirable to identify an experimentally feasible route that enables superconductivity in RP nickelates without pressure or strain.

In this work, we study superconductivity in bulk hole-doped RP nickelate La$_{3-x}$Sr$_x$Ni$_2$O$_7$ under ambient pressure. By combining density functional theory~\cite{PhysRev.136.B864,PhysRev.140.A1133} (DFT), dynamical mean-field theory~\cite{RevModPhys.68.13,RevModPhys.78.865} (DMFT), and random phase approximation~\cite{PhysRevB.31.4403,Graser_2009,RevModPhys.84.1383} (RPA), we solve the linearized gap equation for superconductivity. We find that at small $x$, a Ni-$d_{3z^2-r^2}$-derived hole pocket---known as the $\gamma$ pocket---appears on the Fermi surface. As $x$ increases, the $\gamma$ pocket expands and evolves from a circular to a diamond-like shape. At $x \approx 0.4$, the diamond-shaped $\gamma$ pocket spans nearly half of the Brillouin zone and exhibits almost perfect Fermi surface nesting with the optimal nesting vector $\textbf{Q} = (\pi,\pi)$. This significantly enhances antiferromagnetic spin fluctuations and substantially increases the leading superconducting eigenvalue---several times larger than that of undoped La$_3$Ni$_2$O$_7$ under ambient pressure. With a physically reasonable interaction strength~\cite{327nature,yang2024orbital,PhysRevLett.131.206501}, the leading eigenvalue of the gap equation is sufficiently large to yield experimentally observable superconductivity. Key methodological aspects are outlined below, with complete computational details provided in the Supplementary Information.

\section*{Results}

\begin{figure}[t]
\includegraphics[angle=0,width=\textwidth]{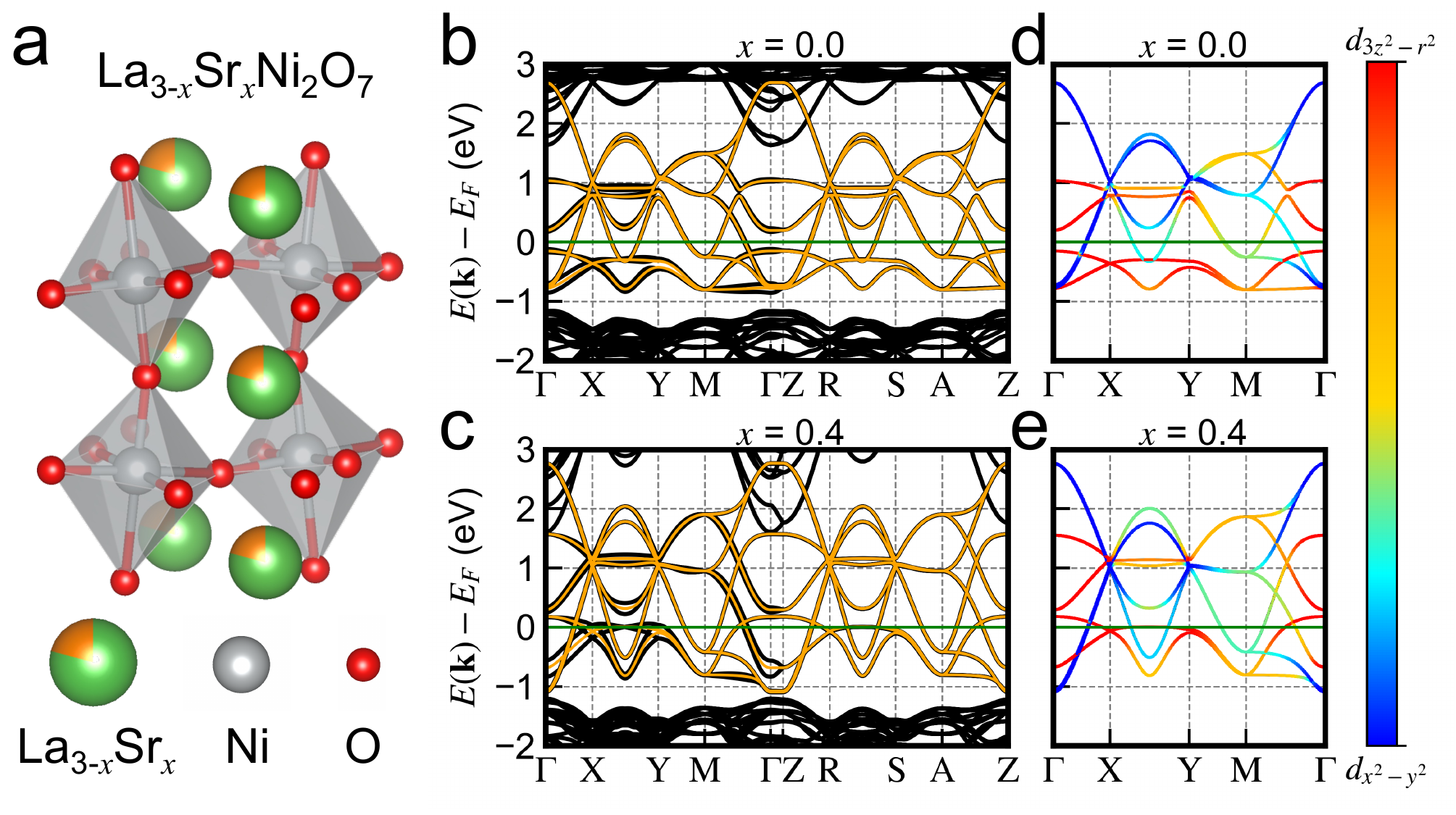}
\caption{\label{fig1} (a) Crystal structure of La$_{3-x}$Sr$_x$Ni$_2$O$_7$ under ambient pressure (space group $Amam$, No.~63). (b)-(c) Comparison of the DFT band structure (black lines) with the band structure reproduced by the bilayer two-orbital model (orange lines) for La$_{3-x}$Sr$_x$Ni$_2$O$_7$ with (b) $x=0.0$ and (c) $x=0.4$. (d)-(e) Orbital projected band structure reproduced by the bilayer two-orbital model for La$_{3-x}$Sr$_x$Ni$_2$O$_7$ with (d) $x=0.0$ and (e) $x=0.4$. In panels (d) and (e), colors denote orbital contributions: red corresponds to Ni-$d_{3z^2-r^2}$ orbital, while blue corresponds to Ni-$d_{x^2-y^2}$ orbital.}
\end{figure}

Figure~\ref{fig1}(a) shows the crystal structure of La$_{3-x}$Sr$_x$Ni$_2$O$_7$ under ambient pressure. It crystallizes in the orthorhombic $Amam$ structure (space group No.~63), characterized by strong oxygen octahedral rotations~\cite{doi:10.1021/jacs.3c13094,doi:10.1021/acs.inorgchem.4c03042}. The structural distortions result in an in-plane $\sqrt{2}\times\sqrt{2}$ expansion of the unit cell. Throughout the calculations, we focus on the folded Brillouin zone corresponding to the $Amam$ unit cell. In panels (b) and (c), the black curves show the DFT band structure of La$_{3-x}$Sr$_x$Ni$_2$O$_7$ for $x=0.0$ and $x=0.4$, respectively. Since the interblock interaction is weak~\cite{PhysRevLett.131.126001,zhang2024structural,xia2025sensitive,PhysRevB.108.L140505,PhysRevB.111.174506,PhysRevLett.131.236002,PhysRevLett.132.106002,PhysRevB.108.L201121}, we employ the widely used bilayer two-orbital model~\cite{PhysRevLett.131.126001,zhang2024structural,xia2025sensitive,PhysRevB.108.L140505,PhysRevB.111.174506,PhysRevLett.131.236002,PhysRevLett.132.106002,PhysRevB.108.L201121} to fit the band structure of La$_{3-x}$Sr$_x$Ni$_2$O$_7$ for each $x$. We show the reproduced band structure by the orange curves in panels (b) and (c). In panels (d) and (e), we show the orbital projections of the band structure that is reproduced by the bilayer two-orbital model for $x=0.0$ and $x=0.4$, respectively. The red (blue) corresponds to Ni-$d_{3z^2-r^2}$ orbital (Ni-$d_{x^2-y^2}$ orbital). We note that at $x=0.0$, the Ni-$d_{3z^2-r^2}$ derived bonding bands are below the Fermi level. This is consistent with the ARPES measurement of La$_3$Ni$_2$O$_7$ under ambient pressure~\cite{yang2024orbital}. With $x$ increasing, the Ni-$d_{3z^2-r^2}$ derived bonding bands move up and cross the Fermi level, leading to an additional hole pocket (known as the $\gamma$ pocket) on the Fermi surface. We highlight that both the shape and size of the $\gamma$ pocket strongly depend on $x$. We find that $x\approx 0.4$ creates the most favorable condition to inducing superconductivity in bulk La$_{3-x}$Sr$_{x}$Ni$_2$O$_7$ under ambient pressure.

\begin{figure}[t]
\includegraphics[angle=0,width=\textwidth]{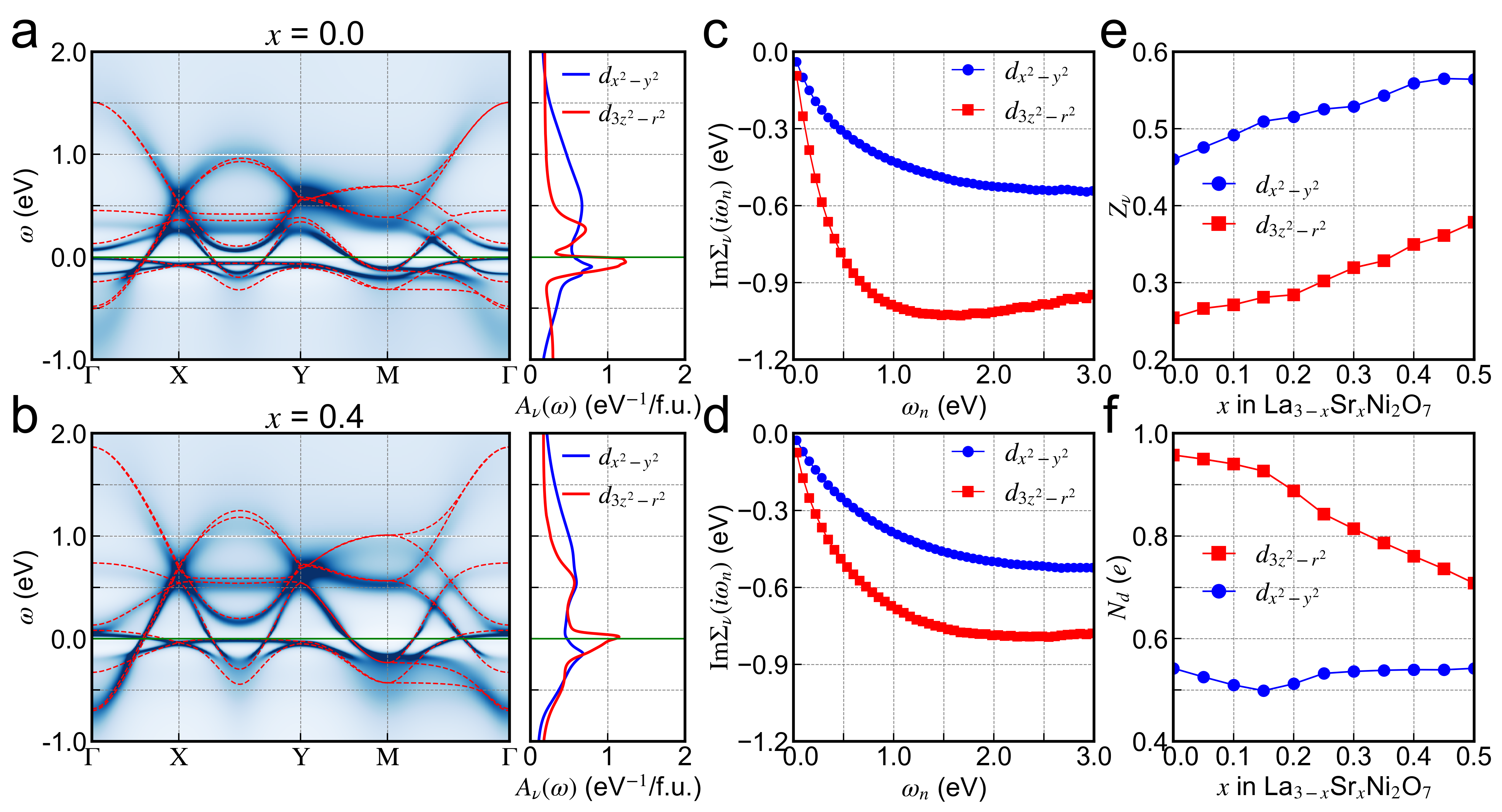}
\caption{\label{fig2} (a)-(b) DMFT-calculated spectral function based on the bilayer two-orbital model for La$_{3-x}$Sr$_x$Ni$_2$O$_7$ with (a) $x=0.0$ and (b) $x=0.4$. Left sub-panel: Momentum-dependent spectral function $A(\textbf{k},\omega)$. The red dashed lines represent the quasiparticle energy dispersion $E(\textbf{k})$ reproduced by the ``dressed'' bilayer two-orbital model. Right sub-panel: Orbital-resolved spectral function $A_{\nu}(\omega)$. (c-d) Orbital-resolved imaginary part of the self-energy $\textrm{Im}\Sigma_{\nu}(i\omega_n)$ of Ni-$d_{3z^2-r^2}$ and Ni-$d_{x^2-y^2}$ orbitals for La$_{3-x}$Sr$_x$Ni$_2$O$_7$ with (c) $x=0.0$ and (d) $x=0.4$. (e) Orbital-resolved quasiparticle weight $Z_{\nu}$ for Ni-$d_{3z^2-r^2}$ and Ni-$d_{x^2-y^2}$ orbitals. (f) Orbital occupancy $N_d$ for Ni-$d_{3z^2-r^2}$ and Ni-$d_{x^2-y^2}$ orbitals. In all panels, we use $U=4$ eV and $J_H=0.5$ eV in the calculations. The red (blue) symbol/curve represents Ni-$d_{3z^2-r^2}$ (Ni-$d_{x^2-y^2}$) orbital.}
\end{figure}

Based on the non-interacting bilayer two-orbital model, we add a Slater-Kanomori interaction~\cite{10.1143/PTP.30.275} and perform DMFT calculations in order to get the quasiparticle dispersion $E(\textbf{k})$ and quasiparticle weight $Z$. Figure~\ref{fig2}(a) and (b) show the interacting electronic structure of the bilayer two-orbital model calculated at $U$ = 4 eV and $J_H$ = 0.5 eV. The left sub-panel is the momentum-dependent spectral function $A(\textbf{k},\omega)$ and the red dashed line in the left sub-panel is the quasiparticle dispersion $E(\textbf{k})$. The right sub-panel is the orbital-resolved spectral function $A_{\nu}(\omega)$, in which the red curve is for Ni-$d_{3z^2-r^2}$ orbital and the blue curve is for Ni-$d_{x^2-y^2}$ orbital. We find that at $x=0.0$ in La$_{3-x}$Sr$_x$Ni$_2$O$_7$, the Ni-$d_{3z^2-r^2}$ spectral function has a peak just below the Fermi level. At $x=0.4$, the peak is shifted to the Fermi level. In panels (c) and (d), we show the imaginary part of the Matsubara self-energy for the Ni-$d_{3z^2-r^2}$ and Ni-$d_{x^2-y^2}$ orbitals. At $x=0.0$, there is a clear orbital-dependent correlation effect~\cite{PhysRevB.108.125105,PhysRevB.108.L201121,PhysRevB.109.L081105,PhysRevB.109.115114,PhysRevB.108.214522}. That is, given the same interaction strength, Ni-$d_{3z^2-r^2}$ orbital has a larger self-energy, which indicates that Ni-$d_{3z^2-r^2}$ orbital is more correlated than Ni-$d_{x^2-y^2}$ orbital. At $x=0.4$, the orbital-dependent correlation effect becomes relatively weaker but Ni-$d_{3z^2-r^2}$ orbital is still the more correlated orbital. Based on the imaginary part of the Matsubara self-energy, we can obtain the quasiparticle weight $Z$. In panels (e) and (f), we show the orbital-resolved quasiparticle weight $Z_{\nu}$ and Ni-$d_{3z^2-r^2}/d_{x^2-y^2}$ occupancy as a function of $x$. We find that for both Ni-$d_{3z^2-r^2}$ and Ni-$d_{x^2-y^2}$ orbitals, the quasiparticle weight $Z_{\nu}$ increases with $x$. The doped holes predominantly reside on Ni-$d_{3z^2-r^2}$ orbital, reducing its occupancy from half-filling, while the occupancy of Ni-$d_{x^2-y^2}$ orbital remains nearly constant and close to quarter-filling with $x$. This indicates that upon increasing $x$, the energy scale for phase coherence is substantially enhanced and exceeds that of pairing~\cite{chen2026unified}. Consequently at a large $x$, the superconducting transition temperature is governed by pairing rather than phase coherence~\cite{Emery1995,chen2026unified}. This motivates our subsequent superconducting gap equation calculations with an effective pairing potential.   

Equipped with the quasiparticle energy $E(\textbf{k})$ and the quasiparticle weight $Z$, we proceed to solve the linearized gap equation using the RPA+DMFT approach~\cite{xia2025nickelate,PhysRevLett.129.077002,PhysRevB.95.174504,PhysRevLett.123.217005,PhysRevLett.123.247001}. The basic building block is the ``dressed'' static susceptibility $\widetilde{\chi}_0(\mathbf{q})_{l_1l_2,l_3l_4}$~\cite{xia2025nickelate}:
\begin{eqnarray}
\label{eq:15}
 \widetilde{\chi}_0(\mathbf{q})_{l_1l_2,l_3l_4}=&&-\sqrt{Z_{l_1}Z_{l_2}Z_{l_3}Z_{l_4}}\times \\\nonumber
 && \frac{1}{N_{\textbf{k}}}\sum_{\textbf{k}}\sum_{p,p'}(A_{\textbf{k}+\textbf{q}})_{l_1 p}(A_{\textbf{k}+\textbf{q}})^{\ast}_{l_3 p}(A_{\textbf{k}})_{l_4 p'}(A_{\textbf{k}})^{\ast}_{ l_2 p'}\frac{n_F[(E_{\textbf{k}+\textbf{q}})_{p}]-n_F[(E_{\textbf{k}})_{p'}]}{(E_{\textbf{k}+\textbf{q}})_{p}-(E_{\textbf{k}})_{p'}}
\end{eqnarray}
where $l_i$ is an orbital index, $p$ is a band index and $A_{\textbf{k}}$ is a transition matrix that connects the orbital basis to the band basis. Using $\widetilde{\chi}_0(\mathbf{q})_{l_1l_2,l_3l_4}$, we construct the charge/spin susceptibility and the pairing potential. The pairing potential is based on a Slater-Kanomori interaction~\cite{10.1143/PTP.30.275} with $U = 4$ eV and $J_H = 0.5$ eV.

\begin{figure}[t]
\includegraphics[angle=0,width=\textwidth]{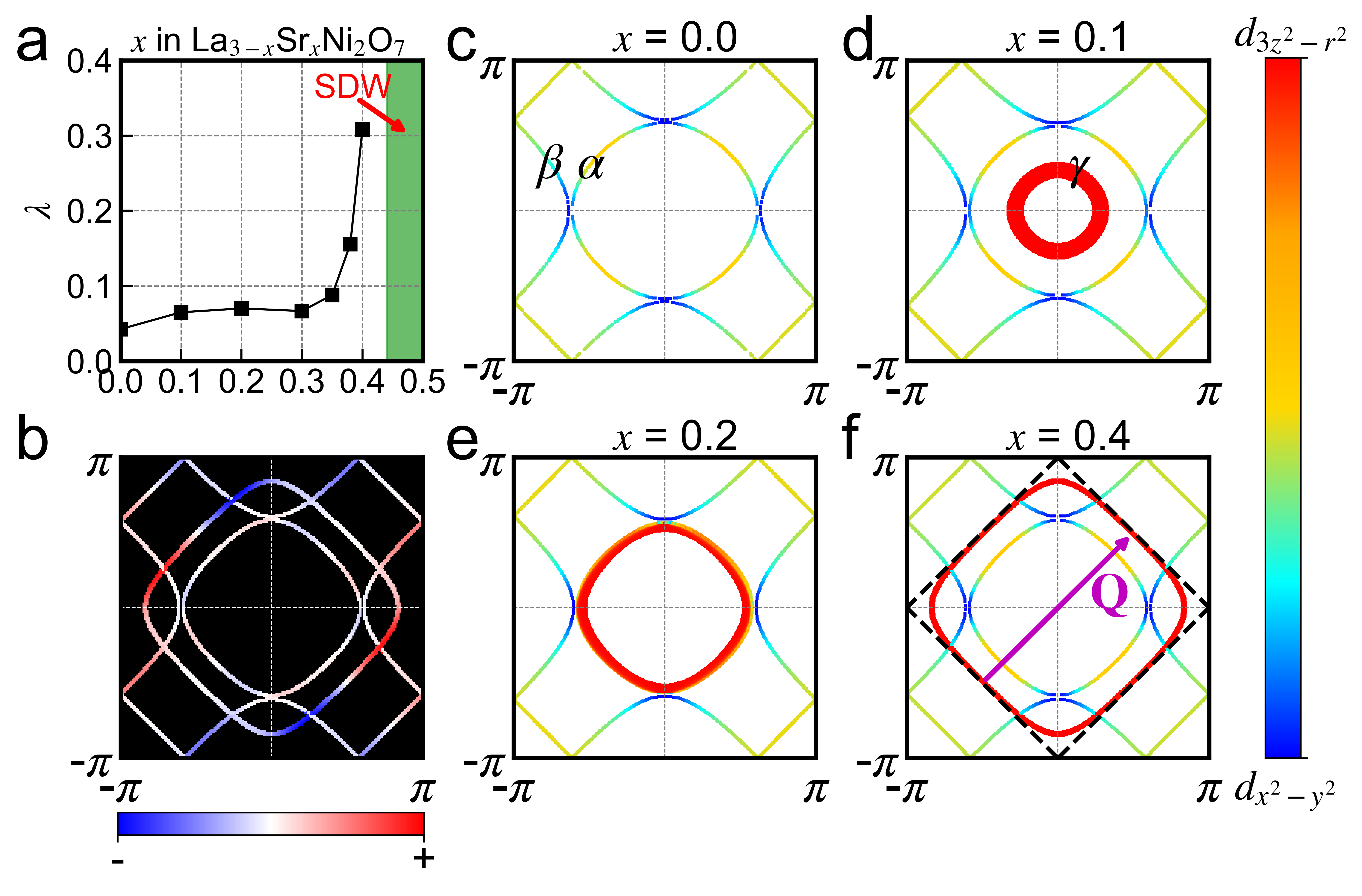}
\caption{\label{fig3} (a) The leading superconducting eigenvalue for La$_{3-x}$Sr$_x$Ni$_2$O$_7$ as a function of $x$. The green shade indicates that the spin-density-wave (SDW) instability is triggered. (b) The gap function of the leading superconducting eigenvalue for La$_{3-x}$Sr$_x$Ni$_2$O$_7$ with $x=0.4$. The gap is nodal. (c)-(f) The orbital-projected Fermi surface for La$_{3-x}$Sr$_x$Ni$_2$O$_7$ with (c) $x=0.0$, (d) $x=0.1$, (e) $x=0.2$, (f) $x=0.4$. In panels (c)-(f), the color indicates the orbital contributions: red is for Ni-$d_{3z^2-r^2}$ orbital and blue is for Ni-$d_{x^2-y^2}$ orbital. In (f), $\mathbf{Q}$ denotes the nesting vector of the Ni-$d_{3z^2-r^2}$-derived $\gamma$ pocket. In all panels, we use $U=4$ eV and $J_H=0.5$ eV in the calculations.}
\end{figure}

Figure~\ref{fig3}(a) shows the leading superconducting eigenvalue $\lambda$ as a function of $x$. $\lambda$ is related to the superconducting transition temperature via $T_c \propto e^{-1/\lambda}$. We find that at $x=0$, $\lambda$ is very small ($<0.05$), consistent with the absence of superconductivity in La$_3$Ni$_2$O$_7$ under ambient pressure~\cite{327nature,zhang2024high,10.1093/nsr/nwaf220}. As $x \approx 0.1-0.2$, $\lambda$ slightly increases but still remains small. However, as $x \approx 0.4$, $\lambda$ substantially rises to above 0.3, which is sufficiently large to result in experimentally observable superconductivity~\cite{9t6n-jqr5,PhysRevB.108.165141,PhysRevB.109.104508,lechermann2023electronic,xia2025sensitive}. When $x$ is further increased, a spin-density-wave instability is triggered. Fig.~\ref{fig3}(b) shows the superconducting gap of La$_{3-x}$Sr$_x$Ni$_2$O$_7$ for $x=0.4$. We find that for the leading eigenvalue, the superconducting gap is nodal with a $d_{x^2-y^2}$-wave symmetry ($A_g$ representation of the point group $D_{2h}$~\cite{RevModPhys.72.969}). We note that because the crystal structure is orthorhombic with the point group $D_{2h}$, the Fermi surface only has $C_2$ rotational symmetry instead of $C_4$ rotational symmetry.

To understand the hole dependence of $\lambda$, in particular the rapid increase of $\lambda$ at $x\approx 0.4$, we show the interacting Fermi surface at a few representative $x$ values in Fig.~\ref{fig3}(c)-(f). At $x=0$, the Fermi surface only has two sheets that are mainly derived from Ni-$d_{x^2-y^2}$ orbital. They are referred to as the $\alpha$ and $\beta$ Fermi sheets. As $x \approx 0.1-0.2$, an additional hole pocket, known as the $\gamma$ pocket, appears and expands around the zone center. We find that when the $\gamma$ pocket just appears on the Fermi surface, or overlaps with the $\alpha$ sheet, its inverse Fermi velocity is large (see Supplementary Note 13), which promotes pairing~\cite{24f4-349n,ushio2025theor,zhang2025compr}. This explains the slight increase of $\lambda$ around $x \approx 0.1-0.2$, known as the ``incipient band'' effect~\cite{PhysRevResearch.2.042032,la3ni2o6,ncbf-9b8m}. As $x$ further increases to 0.4, the $\gamma$ pocket changes its shape from circular to diamond-like. More importantly, the diamond-shaped $\gamma$ pocket spans nearly half of the Brillouin zone. In this configuration, the $\gamma$ pocket exhibits almost perfect Fermi surface nesting with an optimal nesting vector $\textbf{Q}=(\pi,\pi)$, as highlighted by the purple arrow in panel (f). This nesting vector indicates that antiferromagnetic spin fluctuations are significantly enhanced, which leads to the substantial increase of the superconducting eigenvalue $\lambda$ at $x\approx 0.4$. In passing, we note that for completeness, we also perform standard RPA calculations on the hole dependence and find qualitatively similar results (see Supplementary Note 11 for details).

\begin{figure}[t]
\includegraphics[angle=0,width=0.9\textwidth]{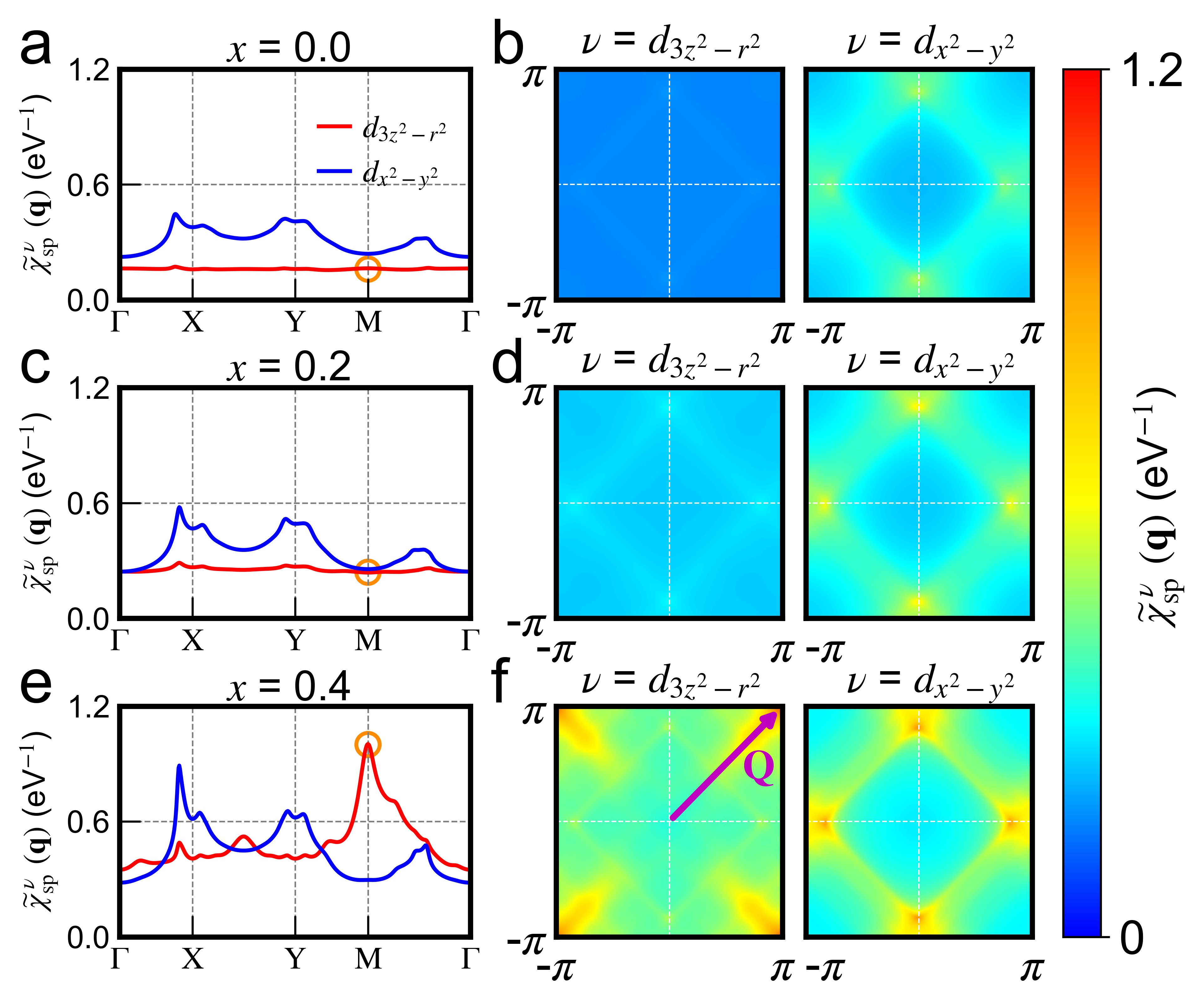}
\caption{\label{fig4} The ``dressed'' orbital-resolved spin susceptibility $\widetilde{\chi}^{\nu}_{\rm{sp}}(\textbf{q})$ for La$_{3-x}$Sr$_x$Ni$_2$O$_7$ with (a)-(b) $x=0.0$, (c)-(d) $x=0.2$, (e)-(f) $x=0.4$. (a),(c),(e) $\widetilde{\chi}^{\nu}_{\rm{sp}}(\textbf{q})$ along the high-symmetry $\textbf{q}$ path. The red (blue) curve represents Ni-$d_{3z^2-r^2}$ (Ni-$d_{x^2-y^2}$) orbital. The orange circle highlights the value of $\widetilde{\chi}^{\nu}_{\rm{sp}}(\textbf{q})$ for $\nu = $Ni-$d_{3z^2-r^2}$ orbital at $M=(\pi,\pi)$ point. (b),(d),(f) $\widetilde{\chi}^{\nu}_{\rm{sp}}(\textbf{q})$ in the two-dimensional Brillouin zone. In (f), the vector $\mathbf{Q}$ represents the nesting vector of Ni-$d_{3z^2-r^2}$ orbital, which is associated with the peak in $\widetilde{\chi}^{\nu}_{\rm{sp}}(\textbf{q})$ at $M=(\pi,\pi)$ point. In all panels, we use $U=4$ eV and $J_H=0.5$ eV in the calculations.}
\end{figure}

To corroborate the preceding analysis, we show in Figure~\ref{fig4} the ``dressed'' orbital-resolved spin susceptibility $\widetilde{\chi}^{\nu}_{\rm{sp}}(\textbf{q})$ for La$_{3-x}$Sr$_x$Ni$_2$O$_7$. $\widetilde{\chi}^{\nu}_{\rm{sp}}(\textbf{q})$ is a measure of spin fluctuations and is calculated based on the quasiparticle energy $E(\textbf{k})$ at $U=4$ eV/$J_H=0.5$ eV. Panels (a) and (b) show $\widetilde{\chi}^{\nu}_{\rm{sp}}(\textbf{q})$ for $x=0.0$ along the high-symmetry $\textbf{q}$ path and in the two-dimensional Brillouin zone. We find that $\widetilde{\chi}^{\nu}_{\rm{sp}}(\textbf{q})$ for Ni-$d_{3z^2-r^2}$ orbital is almost momentum-independent. This is because the Ni-$d_{3z^2-r^2}$-derived $\gamma$ pocket does not appear on the Fermi surface at $x=0.0$. Panels (c) and (d) show $\widetilde{\chi}^{\nu}_{\rm{sp}}(\textbf{q})$ for $x=0.2$. We find that compared to that for $x=0.0$, the average value of $\widetilde{\chi}^{\nu}_{\rm{sp}}(\textbf{q})$ for Ni-$d_{3z^2-r^2}$ orbital is increased, but its momentum dependence is still weak. By contrast, as shown in panels (e) and (f), at $x=0.4$, $\widetilde{\chi}^{\nu}_{\rm{sp}}(\textbf{q})$ for Ni-$d_{3z^2-r^2}$ orbital exhibits a pronounced peak at $M=(\pi,\pi)$ point. This is consistent with the nesting vector $\textbf{Q}=(\pi,\pi)$ of the $\gamma$ pocket, shown in Fig.~\ref{fig3}(f). We note that as $x$ is further increased, the nesting vector deviates from $\textbf{Q}=(\pi,\pi)$, which alone weakens antiferromagnetic spin fluctuations. However, in RPA+DMFT, the ``dressed'' susceptibility $\widetilde{\chi}_0(\mathbf{q})$ defined in Eq.~(\ref{eq:15}) also depends on the quasiparticle weight $Z_{\nu}$, which monotonically increases with $x$. Combining these two factors, we find that with $x$ further increasing, the spin susceptibility $\widetilde{\chi}^{\nu}_{\rm{sp}}(\textbf{q})$ diverges at $M=(\pi, \pi)$ and triggers a spin-density-wave instability, as shown in Fig.~\ref{fig3}(a).


\section*{Discussion and Conclusion}

Before we conclude, we make a few comments.

1) Bulk La$_{3-x}$Sr$_x$Ni$_2$O$_7$ has been synthesized for $x$ up to 0.1. It has been found that as $x$ increases, La$_{3-x}$Sr$_x$Ni$_2$O$_7$ remains metallic and its resistivity drops, but no superconductivity has been observed under ambient pressure~\cite{JIAO20241354504}. Our calculations show that a higher hole concentration of $x\approx 0.4$ creates the most favorable condition for inducing ambient-pressure bulk superconductivity.

2) While a few studies suggest possible magnetic order in La$_3$Ni$_2$O$_7$ under ambient pressure~\cite{chen2024electronic,ni2025spin,PhysRevB.111.184401}, the nuclear magnetic resonance (NMR) and muon-spin rotation/relaxation ($\mu$SR) measurements find that the Ni moments are small~\cite{ZHAO20251239,khasanov2025pressure,PhysRevLett.132.256503}. Furthermore, the ARPES measurement on La$_3$Ni$_2$O$_7$ under ambient pressure reveals that the experimental Fermi surface is in excellent agreement with the non-magnetic DFT calculations~\cite{yang2024orbital}. This lays a solid foundation for using the non-magnetic DFT electronic structure as the starting point in our superconducting calculations.

3) The current study focuses on bulk La$_{3-x}$Sr$_x$Ni$_2$O$_7$ rather than thin films, whose electronic structure is markedly different from the bulk~\cite{hao2025superconductivity,wang2025el}. In bulk La$_3$Ni$_2$O$_7$, the Ni-$d_{3z^2-r^2}$-derived bonding bands lie slightly below the Fermi level. In contrast, in thin films grown on SrLaAlO$_4$ substrates, compressive strain further lowers these bands by approximately 0.3 eV~\cite{wang2025el,PhysRevB.111.115154,85qv-ncxb}. Consequently, we find that for a wide range of hole doping ($x$ from 0.0 to 1.0), the Fermi surface of the compressively strained films lacks a large diamond-shaped $\gamma$ pocket with strong nesting. In addition, hole doping substantially reduces the $\alpha$/$\beta$ Fermi sheets, leading to a pronounced reconstruction of the Fermi surface in La$_{3-x}$Sr$_x$Ni$_2$O$_7$ thin films. See Supplementary Note 16 for details. These results suggest that bulk and thin-film La$_{3-x}$Sr$_x$Ni$_2$O$_7$ exhibit distinct behaviors upon hole doping.

4) Under high pressure ($> 10$ GPa) or compressive strain ($\approx 2\%$), La$_3$Ni$_2$O$_7$ crystallizes in a high-symmetry tetragonal $I4/mmm$ structure. By contrast, under ambient pressure and at $x \approx 0.4$, La$_{3-x}$Sr$_x$Ni$_2$O$_7$ crystallizes in the low-symmetry orthorhombic $Amam$ structure. This indicates that spin fluctuations, rather than the high-symmetry structure, play a more important role in enabling superconductivity in RP nickelates, consistent with the recent experimental study~\cite{shi2025spin}. We argue that although a high-symmetry tetragonal structure may facilitate pairing, it is not a necessary condition for superconductivity.

In conclusion, using the realistic electronic structure and solving the superconducting gap equation, we find a substantially increased superconducting eigenvalue in bulk La$_{3-x}$Sr$_x$Ni$_2$O$_7$ at $x \approx 0.4$ under ambient pressure, yielding observable superconductivity. The underlying mechanism is that, with hole doping, the Ni-$d_{3z^2-r^2}$-derived bonding bands cross the Fermi level and induce a $\gamma$ hole pocket on the Fermi surface. The hole concentration $x$ serves as a tuning parameter that controls both the size and shape of the $\gamma$ pocket. As $x$ increases, the $\gamma$ pocket expands and evolves from a circular to a diamond-like shape. At $x \approx 0.4$, the diamond-shaped $\gamma$ pocket spans nearly half of the Brillouin zone and exhibits almost perfect Fermi surface nesting with an optimal nesting vector $\textbf{Q} = (\pi,\pi)$. This substantially enhances antiferromagnetic spin fluctuations and, consequently, induces unconventional superconductivity. Our work identifies a simple yet robust mechanism and a feasible experimental route to inducing the elusive bulk superconductivity in Ruddlesden-Popper nickelates without pressure or strain.


\begin{acknowledgments}
We are grateful to Fan Yang and Huiqiu Yuan for insightful discussions. This project was financially supported by the National Natural Science Foundation of China under project number 12374064 and 12434002, Science and Technology Commission of Shanghai Municipality under grant number 23ZR1445400. C.X. was supported by the National Natural Science Foundation of China under project number 12404082. NYU High-Performance-Computing (HPC) provides computational resources.
\end{acknowledgments}

\bibliographystyle{apsrev4-2}
\bibliography{main}

@article{ncbf-9b8m,
  title = {Superconductivity Governed by Janus-Faced Fermiology in Strained Bilayer Nickelates},
  author = {Ryee, Siheon and Witt, Niklas and Sangiovanni, Giorgio and Wehling, Tim O.},
  journal = {Phys. Rev. Lett.},
  volume = {135},
  issue = {23},
  pages = {236003},
  numpages = {8},
  year = {2025},
  month = {Dec}
}

@article{PhysRevResearch.2.042032,
  title = {Designing nickelate superconductors with ${d}^{8}$ configuration exploiting mixed-anion strategy},
  author = {Kitamine, Naoya and Ochi, Masayuki and Kuroki, Kazuhiko},
  journal = {Phys. Rev. Res.},
  volume = {2},
  issue = {4},
  pages = {042032},
  numpages = {7},
  year = {2020},
  month = {Nov}
}

@article{la3ni2o6,
      title={Theoretical proposal of superconductivity in hole-doped reduced bilayer nickelate La3Ni2O6: a manifestation of orbital-space bilayer model with incipient bands}, 
      author={Shu Kamiyama and Reo Kohno and Yuto Hoshi and Kensei Ushio and Daiki Nakaoka and Hirofumi Sakakibara and Kazuhiko Kuroki},
      year={2026},
      journal={arXiv:2603.11771}
}

@article{PhysRevLett.123.217005,
  title = {Superconducting Symmetries of ${\mathrm{Sr}}_{2}{\mathrm{RuO}}_{4}$ from First-Principles Electronic Structure},
  author = {Gingras, O. and Nourafkan, R. and Tremblay, A.-M. S. and C\^ot\'e, M.},
  journal = {Phys. Rev. Lett.},
  volume = {123},
  issue = {21},
  pages = {217005},
  numpages = {7},
  year = {2019}
}

@article{lechermann2023electronic,
  title = {Electronic correlations and superconducting instability in $\mathrm{La_3Ni_2O_7}$ under high pressure},
  author = {Lechermann, Frank and Gondolf, Jannik and B\"otzel, Steffen and Eremin, Ilya M.},
  journal = {Phys. Rev. B},
  volume = {108},
  issue = {20},
  pages = {L201121},
  numpages = {6},
  year = {2023},
  month = {Nov},
  publisher = {American Physical Society}
}

@article{Graser_2009,
year = {2009},
month = {feb},
volume = {11},
number = {2},
pages = {025016},
author = {S Graser and T A Maier and P J Hirschfeld and D J Scalapino},
title = {Near-degeneracy of several pairing channels in multiorbital models for the Fe pnictides},
journal = {New J. Phys.}
}

@article{RevModPhys.84.1383,
  title = {A common thread: The pairing interaction for unconventional superconductors},
  author = {Scalapino, D. J.},
  journal = {Rev. Mod. Phys.},
  volume = {84},
  issue = {4},
  pages = {1383--1417},
  numpages = {0},
  year = {2012},
  month = {Oct},
  publisher = {American Physical Society}
}

@article{PhysRevB.31.4403,
  title = {Two-dimensional Hubbard model: Numerical simulation study},
  author = {Hirsch, J. E.},
  journal = {Phys. Rev. B},
  volume = {31},
  issue = {7},
  pages = {4403--4419},
  numpages = {0},
  year = {1985},
  month = {Apr},
  publisher = {American Physical Society}
}

@article{RevModPhys.68.13,
  title = {Dynamical mean-field theory of strongly correlated fermion systems and the limit of infinite dimensions},
  author = {Georges, Antoine and Kotliar, Gabriel and Krauth, Werner and Rozenberg, Marcelo J.},
  journal = {Rev. Mod. Phys.},
  volume = {68},
  issue = {1},
  pages = {13--125},
  numpages = {0},
  year = {1996},
  month = {Jan},
  publisher = {American Physical Society}
}

@article{PhysRevB.95.174504,
  title = {Orbital selective pairing and gap structures of iron-based superconductors},
  author = {Kreisel, Andreas and Andersen, Brian M. and Sprau, P. O. and Kostin, A. and Davis, J. C. S\'eamus and Hirschfeld, P. J.},
  journal = {Phys. Rev. B},
  volume = {95},
  issue = {17},
  pages = {174504},
  numpages = {12},
  year = {2017},
  month = {May},
  publisher = {American Physical Society}
}

@article{PhysRevLett.129.077002,
  title = {Superconducting Instabilities in Strongly Correlated Infinite-Layer Nickelates},
  author = {Kreisel, Andreas and Andersen, Brian M. and R{{\O{}}}mer, Astrid T. and Eremin, Ilya M. and Lechermann, Frank},
  journal = {Phys. Rev. Lett.},
  volume = {129},
  issue = {7},
  pages = {077002},
  numpages = {7},
  year = {2022},
  month = {Aug},
  publisher = {American Physical Society}
}

@article{PhysRev.136.B864,
  title = {Inhomogeneous Electron Gas},
  author = {Hohenberg, P. and Kohn, W.},
  journal = {Phys. Rev.},
  volume = {136},
  issue = {3B},
  pages = {B864--B871},
  numpages = {0},
  year = {1964},
  month = {Nov},
  publisher = {American Physical Society}
}

@article{PhysRev.140.A1133,
  title = {Self-Consistent Equations Including Exchange and Correlation Effects},
  author = {Kohn, W. and Sham, L. J.},
  journal = {Phys. Rev.},
  volume = {140},
  issue = {4A},
  pages = {A1133--A1138},
  numpages = {0},
  year = {1965},
  month = {Nov},
  publisher = {American Physical Society}
}

@article{PhysRevLett.123.247001,
  title = {Knight Shift and Leading Superconducting Instability from Spin Fluctuations in ${\mathrm{Sr}}_{2}{\mathrm{RuO}}_{4}$},
  author = {R{{\O{}}}mer, A. T. and Scherer, D. D. and Eremin, I. M. and Hirschfeld, P. J. and Andersen, B. M.},
  journal = {Phys. Rev. Lett.},
  volume = {123},
  issue = {24}, 
  pages = {247001},
  numpages = {6},
  year = {2019},
  month = {Dec},
  publisher = {American Physical Society} 
}

@article{RevModPhys.78.865,
  title = {Electronic structure calculations with dynamical mean-field theory},
  author = {Kotliar, G. and Savrasov, S. Y. and Haule, K. and Oudovenko, V. S. and Parcollet, O. and Marianetti, C. A.},
  journal = {Rev. Mod. Phys.},
  volume = {78},
  issue = {3},
  pages = {865--951},
  numpages = {0},
  year = {2006},
  month = {Aug},
  publisher = {American Physical Society} 
}

@article{PhysRevB.108.L201121,
  title = {Electronic correlations and superconducting instability in ${\mathrm{La}}_{3}{\mathrm{Ni}}_{2}{\mathrm{O}}_{7}$ under high pressure},
  author = {Lechermann, Frank and Gondolf, Jannik and B\"otzel, Steffen and Eremin, Ilya M.},
  journal = {Phys. Rev. B},
  volume = {108},
  issue = {20},
  pages = {L201121},
  numpages = {6},
  year = {2023},
  month = {Nov},
  publisher = {American Physical Society}
}

@article{327nature,
  title={Signatures of superconductivity near 80 \uppercase{K} in a nickelate under high pressure},
  author={Sun, Hualei and Huo, Mengwu and Hu, Xunwu and Li, Jingyuan and Liu, Zengjia and Han, Yifeng and Tang, Lingyun and Mao, Zhongquan and Yang, Pengtao and Wang, Bosen and Cheng, Jinguang and Yao, Dao-Xin and Zhang, Guang-Ming and Wang, Meng},
  journal={Nature},
  volume={621},
  number={7979},
  pages={493--498},
  year={2023},
  publisher={Nature Publishing Group UK London}
}

@article{zhang2024high,
  title={High-temperature superconductivity with zero resistance and strange-metal behaviour in La3Ni2O7- $\delta$},
  author={Zhang, Yanan and Su, Dajun and Huang, Yanen and Shan, Zhaoyang and Sun, Hualei and Huo, Mengwu and Ye, Kaixin and Zhang, Jiawen and Yang, Zihan and Xu, Yongkang and Su, Yi and Li, Rui and Smidman, Michael and Wang, Meng and Jiao, Lin and Yuan, Huiqiu},
  journal={Nat. Phys.},
  volume={20},
  number={8},
  pages={1269--1273},
  year={2024},
  publisher={Nature Publishing Group UK London}
}

@article{wang2024bulk,
  title={Bulk high-temperature superconductivity in pressurized tetragonal La2PrNi2O7},
  author={Wang, Ningning and Wang, Gang and Shen, Xiaoling and Hou, Jun and Luo, Jun and Ma, Xiaoping and Yang, Huaixin and Shi, Lifen and Dou, Jie and Feng, Jie and Yang, Jie and Shi, Yunqing and Ren, Zhian and Ma, Hanming and Yang, Pengtao and Liu, Ziyi and Liu, Yue and Zhang, Hua and Dong, Xiaoli and Wang, Yuxin and Jiang, Kun and Hu, Jiangping and Nagasaki, Shoko and Kitagawa, Kentaro and Calder, Stuart and Yan, Jiaqiang and Sun, Jianping and Wang, Bosen and Zhou, Rui and Uwatoko, Yoshiya and Cheng, Jinguang},
  journal={Nature},
  volume={634},
  number={8034},
  pages={579--584},
  year={2024},
  publisher={Nature Publishing Group UK London}
}

@article{li2026bulk,
  title={Bulk superconductivity up to 96 K in pressurized nickelate single crystals},
  author={Li, Feiyu and Xing, Zhenfang and Peng, Di and Dou, Jie and Guo, Ning and Ma, Liang and Zhang, Yulin and Wang, Lingzhen and Luo, Jun and Yang, Jie and Zhang, Jian and Chang, Tieyan and Chen, Yu-Sheng and Cai, Weizhao and Cheng, Jinguang and Wang, Yuzhu and Liu, Yuxin and Luo, Tao and Hirao, Naohisa and Matsuoka, Takahiro and Kadobayashi, Hirokazu and Zeng, Zhidan and Zheng, Qiang and Zhou, Rui and Zeng, Qiaoshi and Tao, Xutang and Zhang, Junjie},
  journal={Nature},
  volume={649},
  number={8098},
  pages={871--878},
  year={2026},
  publisher={Nature Publishing Group}
}

@article{ko2025signatures,
  title={Signatures of ambient pressure superconductivity in thin film La3Ni2O7},
  author={Ko, Eun Kyo and Yu, Yijun and Liu, Yidi and Bhatt, Lopa and Li, Jiarui and Thampy, Vivek and Kuo, Cheng-Tai and Wang, Bai Yang and Lee, Yonghun and Lee, Kyuho and Lee, Jun-Sik and Goodge, Berit H. and Muller, David A. and Hwang, Harold Y.},
  journal={Nature},
  volume={638},
  number={8052},
  pages={935--940},
  year={2025},
  publisher={Nature Publishing Group UK London}
}

@article{zhou2025ambient,
  title={Ambient-pressure superconductivity onset above 40 K in (La, Pr) 3Ni2O7 films},
  author={Zhou, Guangdi and Lv, Wei and Wang, Heng and Nie, Zihao and Chen, Yaqi and Li, Yueying and Huang, Haoliang and Chen, Wei-Qiang and Sun, Yu-Jie and Xue, Qi-Kun and Chen, Zhuoyu},
  journal={Nature},
  volume={640},
  number={8059},
  pages={641--646},
  year={2025},
  publisher={Nature Publishing Group UK London}
}

@article{hao2025superconductivity,
  title={Superconductivity in Sr-doped La3Ni2O7 thin films},
  author={Hao, Bo and Wang, Maosen and Sun, Wenjie and Yang, Yang and Mao, Zhangwen and Yan, Shengjun and Sun, Haoying and Zhang, Hongyi and Han, Lu and Gu, Zhengbin and Zhou, Jian and Ji, Dianxiang and Nie, Yuefeng},
  journal={Nat. Mater.},
  volume={24},
  number={11},
  pages={1756--1762},
  year={2025},
  publisher={Nature Publishing Group UK London}
}

@article{qrkk-l2ng,
  title = {Superconducting Dome in ${\mathrm{La}}_{3\ensuremath{-}x}{\mathrm{Sr}}_{x}{\mathrm{Ni}}_{2}{\mathrm{O}}_{7\ensuremath{-}\ensuremath{\delta}}$ Thin Films},
  author = {Wang, Maosen and Hao, Bo and Sun, Wenjie and Yan, Shengjun and Sun, Shengwang and Zhang, Hongyi and Gu, Zhengbin and Nie, Yuefeng},
  journal = {Phys. Rev. Lett.},
  volume = {136},
  issue = {6},
  pages = {066002},
  numpages = {8},
  year = {2026},
  month = {Feb},
  publisher = {American Physical Society}
}

@article{liu2025superconductivity,
  title={Superconductivity and normal-state transport in compressively strained La2PrNi2O7 thin films},
  author={Liu, Yidi and Ko, Eun Kyo and Tarn, Yaoju and Bhatt, Lopa and Li, Jiarui and Thampy, Vivek and Goodge, Berit H and Muller, David A and Raghu, Srinivas and Yu, Yijun and Hwang, Harold Y.},
  journal={Nat. Mater.},
  volume={24},
  number={8},
  pages={1221--1227},
  year={2025},
  publisher={Nature Publishing Group UK London}
}

@article{bednorz1986possible,
  title={Possible high T c superconductivity in the Ba- La- Cu- O system},
  author={Bednorz, J George and M{\"u}ller, K Alex},
  journal={Z. Physik B - Condensed Matter},
  volume={64},
  number={2},
  pages={189--193},
  year={1986},
  publisher={Springer}
}

@article{PhysRevLett.58.908,
  title = {Superconductivity at 93 K in a new mixed-phase Y-Ba-Cu-O compound system at ambient pressure},
  author = {Wu, M. K. and Ashburn, J. R. and Torng, C. J. and Hor, P. H. and Meng, R. L. and Gao, L. and Huang, Z. J. and Wang, Y. Q. and Chu, C. W.},
  journal = {Phys. Rev. Lett.},
  volume = {58},
  issue = {9},
  pages = {908--910},
  numpages = {0},
  year = {1987},
  month = {Mar},
  publisher = {American Physical Society}
}

@article{schilling1993superconductivity,
  title={Superconductivity above 130 k in the hg--ba--ca--cu--o system},
  author={Schilling, Andreas and Cantoni, M and Guo, JD and Ott, HR},
  journal={Nature},
  volume={363},
  number={6424},
  pages={56--58},
  year={1993}
}

@article{PhysRevLett.101.107006,
  title = {Superconductivity at 38 K in the Iron Arsenide $({\mathrm{Ba}}_{1\ensuremath{-}x}{\mathrm{K}}_{x}){\mathrm{Fe}}_{2}{\mathrm{As}}_{2}$},
  author = {Rotter, Marianne and Tegel, Marcus and Johrendt, Dirk},
  journal = {Phys. Rev. Lett.},
  volume = {101},
  issue = {10},
  pages = {107006},
  numpages = {4},
  year = {2008},
  month = {Sep},
  publisher = {American Physical Society}
}

@article{doi:10.1021/ja800073m,
author = {Kamihara, Yoichi and Watanabe, Takumi and Hirano, Masahiro and Hosono, Hideo},
title = {Iron-Based Layered Superconductor La[O1-xFx]FeAs (x = 0.05−0.12) with Tc = 26 K},
journal = {J. Am. Chem. Soc.},
volume = {130},
number = {11},
pages = {3296-3297},
year = {2008}
}

@article{doi:10.7566/JPSJ.93.053702,
author = {Kakoi ,Masataka and Oi ,Takashi and Ohshita ,Yujiro and Yashima ,Mitsuharu and Kuroki ,Kazuhiko and Kato ,Takeru and Takahashi ,Hidefumi and Ishiwata ,Shintaro and Adachi ,Yoshinobu and Hatada ,Naoyuki and Uda ,Tetsuya and Mukuda ,Hidekazu},
title = {Multiband Metallic Ground State in Multilayered Nickelates La3Ni2O7 and La4Ni3O10 Probed by 139La-NMR at Ambient Pressure},
journal = {J. Phys. Soc. Jpn.},
volume = {93},
number = {5},
pages = {053702},
year = {2024}
}

@article{XIE20243221,
title = {Strong interlayer magnetic exchange coupling in La3Ni2O7−δ revealed by inelastic neutron scattering},
journal = {Sci. Bull.},
volume = {69},
number = {20},
pages = {3221-3227},
year = {2024},
issn = {2095-9273},
author = {Tao Xie and Mengwu Huo and Xiaosheng Ni and Feiran Shen and Xing Huang and Hualei Sun and Helen C. Walker and Devashibhai Adroja and Dehong Yu and Bing Shen and Lunhua He and Kun Cao and Meng Wang}
}

@article{PhysRevLett.132.146002,
  title = {Interlayer-Coupling-Driven High-Temperature Superconductivity in ${\mathrm{La}}_{3}{\mathrm{Ni}}_{2}{\mathrm{O}}_{7}$ under Pressure},
  author = {Lu, Chen and Pan, Zhiming and Yang, Fan and Wu, Congjun},
  journal = {Phys. Rev. Lett.},
  volume = {132},
  issue = {14},
  pages = {146002},
  numpages = {6},
  year = {2024},
  month = {Apr}
}

@article{shi2025spin,
  title={Spin density wave rather than tetragonal structure is prerequisite for superconductivity in La3Ni2O7-$\delta$},
  author={Shi, Mengzhu and Peng, Di and Li, Yikang and Yang, Shaohua and Xing, Zhenfang and Wang, Yuzhu and Fan, Kaibao and Li, Houpu and Wu, Rongqi and Ge, Binghui and Zeng, Zhidan and Zeng, Qiaoshi and Ying, Jianjun and Wu, Tao and Chen, Xianhui},
  journal={Nat. Commun.},
  volume={16},
  number={1},
  pages={9141},
  year={2025},
  publisher={Nature Publishing Group UK London}
}

@article{Wang_2024,
year = {2024},
month = {jul},
publisher = {Chinese Physical Society and IOP Publishing Ltd},
volume = {41},
number = {7},
pages = {077402},
author = {Wang, Meng and Wen, Hai-Hu and Wu, Tao and Yao, Dao-Xin and Xiang, Tao},
title = {Normal and Superconducting Properties of La3Ni2O7},
journal = {Chin. Phys. Lett.}
}

@Article{cpl_41_1_017401,
title = {Signature of Superconductivity in Pressurized La<sub>4</sub>Ni<sub>3</sub>O<sub>10</sub>},
journal = {Chin. Phys. Lett.},
volume = {41},
number = {1},
pages = {017401},
year = {2024},
author = {Qing Li and Ying-Jie Zhang and Zhe-Ning Xiang and Yuhang Zhang and Xiyu Zhu and Hai-Hu Wen}
}

@article{Zhu2024,
year = {2024},
volume = {631},
pages = {531--536},
author = {Zhu, Yinghao and Peng, Di and Zhang, Enkang and Pan, Bingying and Chen, Xu and Chen, Lixing and Ren, Huifen and Liu, Feiyang and Hao, Yiqing and Li, Nana and Xing, Zhenfang and Lan, Fujun and Han, Jiyuan and Wang, Junjie and Jia, Donghan and Wo, Hongliang and Gu, Yiqing and Gu, Yimeng and Ji, Li and Wang, Wenbin and Gou, Huiyang and Shen, Yao and Ying, Tianping and Chen, Xiaolong and Yang, Wenge and Cao, Huibo and Zheng, Changlin and Zeng, Qiaoshi and Guo, Jian-Gang and Zhao, Jun},
title = {Superconductivity in pressurized trilayer La$_4$Ni$_3$O$_{10-\delta}$ single crystals},
journal = {Nature}
}

@article{PhysRevX.15.021005,
  title = {Superconductivity in Trilayer Nickelate ${\mathrm{La}}_{4}{\mathrm{Ni}}_{3}{\mathrm{O}}_{10}$ under Pressure},
  author = {Zhang, Mingxin and Pei, Cuiying and Peng, Di and Du, Xian and Hu, Weixiong and Cao, Yantao and Wang, Qi and Wu, Juefei and Li, Yidian and Liu, Huanyu and Wen, Chenhaoping and Song, Jing and Zhao, Yi and Li, Changhua and Cao, Weizheng and Zhu, Shihao and Zhang, Qing and Yu, Na and Cheng, Peihong and Zhang, Lili and Li, Zhiwei and Zhao, Jinkui and Chen, Yulin and Jin, Changqing and Guo, Hanjie and Wu, Congjun and Yang, Fan and Zeng, Qiaoshi and Yan, Shichao and Yang, Lexian and Qi, Yanpeng},
  journal = {Phys. Rev. X},
  volume = {15},
  issue = {2},
  pages = {021005},
  numpages = {11},
  year = {2025},
  month = {Apr},
  publisher = {American Physical Society}
}

@article{liu2023evidence,
  title={Evidence for charge and spin density waves in single crystals of La3Ni2O7 and La3Ni2O6},
  author={Liu, Zengjia and Sun, Hualei and Huo, Mengwu and Ma, Xiaoyan and Ji, Yi and Yi, Enkui and Li, Lisi and Liu, Hui and Yu, Jia and Zhang, Ziyou and Chen, Zhiqiang and Liang, Feixiang and Dong, Hongliang and Guo, Hanjie and Zhong, Dingyong and Shen, Bing and Li, Shiliang and Wang, Meng},
  journal={Sci. China Phys. Mech. Astron.},
  volume={66},
  number={1},
  pages={217411},
  year={2023},
  publisher={Springer}
}

@article{PhysRevLett.132.256503,
  title = {Evidence of Spin Density Waves in ${\mathrm{La}}_{3}{\mathrm{Ni}}_{2}{\mathrm{O}}_{7\ensuremath{-}\ensuremath{\delta}}$},
  author = {Chen, Kaiwen and Liu, Xiangqi and Jiao, Jiachen and Zou, Muyuan and Jiang, Chengyu and Li, Xin and Luo, Yixuan and Wu, Qiong and Zhang, Ningyuan and Guo, Yanfeng and Shu, Lei},
  journal = {Phys. Rev. Lett.},
  volume = {132},
  issue = {25},
  pages = {256503},
  numpages = {7},
  year = {2024},
  month = {Jun},
  publisher = {American Physical Society}
}

@article{chen2024electronic,
  title={Electronic and magnetic excitations in La3Ni2O7},
  author={Chen, Xiaoyang and Choi, Jaewon and Jiang, Zhicheng and Mei, Jiong and Jiang, Kun and Li, Jie and Agrestini, Stefano and Garcia-Fernandez, Mirian and Sun, Hualei and Huang, Xing and Shen, Dawei and Wang, Meng and Hu, Jiangping and Lu, Yi and Zhou, Ke-Jin and Feng, Donglai},
  journal={Nat. Commun.},
  volume={15},
  number={1},
  pages={9597},
  year={2024},
  publisher={Nature Publishing Group UK London}
}

@article{khasanov2025pressure,
  title={Pressure-enhanced splitting of density wave transitions in La3Ni2O7--$\delta$},
  author={Khasanov, Rustem and Hicken, Thomas J and Gawryluk, Dariusz J and Sazgari, Vahid and Plokhikh, Igor and Sorel, Lo{\"\i}c Pierre and Bartkowiak, Marek and B{\"o}tzel, Steffen and Lechermann, Frank and Eremin, Ilya M and Luetkens, Hubertus and Guguchia, Zurab},
  journal={Nat. Phys.},
  volume={21},
  number={3},
  pages={430--436},
  year={2025},
  publisher={Nature Publishing Group UK London}
}

@article{PhysRevB.110.205122,
  title = {Electronic and magnetic structures of bilayer ${\text{La}}_{3}{\text{Ni}}_{2}{\text{O}}_{7}$ at ambient pressure},
  author = {Wang, Yuxin and Jiang, Kun and Wang, Ziqiang and Zhang, Fu-Chun and Hu, Jiangping},
  journal = {Phys. Rev. B},
  volume = {110},
  issue = {20},
  pages = {205122},
  numpages = {7},
  year = {2024},
  month = {Nov},
  publisher = {American Physical Society}
}

@article{ni2025spin,
  title={Spin density wave in the bilayered nickelate La3Ni2O7- $\delta$ at ambient pressure},
  author={Ni, Xiao-Sheng and Ji, Yuyang and He, Lixin and Xie, Tao and Yao, Dao-Xin and Wang, Meng and Cao, Kun},
  journal={npj Quantum Mater.},
  volume={10},
  number={1},
  pages={17},
  year={2025},
  publisher={Nature Publishing Group UK London}
}

@article{PhysRevB.111.184401,
  title = {Spin-charge-orbital order in nickelate superconductors},
  author = {Zhang, Binhua and Xu, Changsong and Xiang, Hongjun},
  journal = {Phys. Rev. B},
  volume = {111},
  issue = {18},
  pages = {184401},
  numpages = {6},
  year = {2025},
  month = {May},
  publisher = {American Physical Society}
}

@article{PhysRevLett.132.106002,
  title = {Possible High ${T}_{c}$ Superconductivity in ${\mathrm{La}}_{3}{\mathrm{Ni}}_{2}{\mathrm{O}}_{7}$ under High Pressure through Manifestation of a Nearly Half-Filled Bilayer Hubbard Model},
  author = {Sakakibara, Hirofumi and Kitamine, Naoya and Ochi, Masayuki and Kuroki, Kazuhiko},
  journal = {Phys. Rev. Lett.},
  volume = {132},
  issue = {10},
  pages = {106002},
  numpages = {6},
  year = {2024},
  month = {Mar},
  publisher = {American Physical Society}
}

@article{PhysRevLett.131.236002,
  title = {${s}^{\ifmmode\pm\else\textpm\fi{}}$-Wave Pairing and the Destructive Role of Apical-Oxygen Deficiencies in ${\mathrm{La}}_{3}{\mathrm{Ni}}_{2}{\mathrm{O}}_{7}$ under Pressure},
  author = {Liu, Yu-Bo and Mei, Jia-Wei and Ye, Fei and Chen, Wei-Qiang and Yang, Fan},
  journal = {Phys. Rev. Lett.},
  volume = {131},
  issue = {23},
  pages = {236002},
  numpages = {6},
  year = {2023},
  month = {Dec},
  publisher = {American Physical Society}
}

@article{PhysRevLett.131.206501,
  title = {Correlated Electronic Structure of ${\mathrm{La}}_{3}{\text{Ni}}_{2}{\mathrm{O}}_{7}$ under Pressure},
  author = {Christiansson, Viktor and Petocchi, Francesco and Werner, Philipp},
  journal = {Phys. Rev. Lett.},
  volume = {131},
  issue = {20},
  pages = {206501},
  numpages = {6},
  year = {2023},
  month = {Nov},
  publisher = {American Physical Society}
}

@article{PhysRevLett.131.126001,
  title = {Bilayer Two-Orbital Model of $\mathrm{L}{\mathrm{a}}_{3}\mathrm{N}{\mathrm{i}}_{2}{\mathrm{O}}_{7}$ under Pressure},
  author = {Luo, Zhihui and Hu, Xunwu and Wang, Meng and W\'u, W\'ei and Yao, Dao-Xin},
  journal = {Phys. Rev. Lett.},
  volume = {131},
  issue = {12},
  pages = {126001},
  numpages = {6},
  year = {2023},
  month = {Sep},
  publisher = {American Physical Society}
}

@article{yang2024orbital,
  title={Orbital-dependent electron correlation in double-layer nickelate La3Ni2O7},
  author={Yang, Jiangang and Sun, Hualei and Hu, Xunwu and Xie, Yuyang and Miao, Taimin and Luo, Hailan and Chen, Hao and Liang, Bo and Zhu, Wenpei and Qu, Gexing and Chen, Cui-Qun and Huo, Mengwu and Huang, Yaobo and Zhang, Shenjin and Zhang, Fengfeng and Yang, Feng and Wang, Zhimin and Peng, Qinjun and Mao, Hanqing and Liu, Guodong and Xu, Zuyan and Qian, Tian and Yao, Dao-Xin and Wang, Meng and Zhao, Lin and Zhou, X. J.},
  journal={Nat. Commun.},
  volume={15},
  number={1},
  pages={4373},
  year={2024},
  publisher={Nature Publishing Group UK London}
}

@article{zhang2024structural,
  title={Structural phase transition, s$\pm$-wave pairing, and magnetic stripe order in bilayered superconductor La3Ni2O7 under pressure},
  author={Zhang, Yang and Lin, Ling-Fang and Moreo, Adriana and Maier, Thomas A and Dagotto, Elbio},
  journal={Nat. Commun.},
  volume={15},
  number={1},
  pages={2470},
  year={2024},
  publisher={Nature Publishing Group UK London}
}

@article{xia2025sensitive,
  title={Sensitive dependence of pairing symmetry on Ni-eg crystal field splitting in the nickelate superconductor La3Ni2O7},
  author={Xia, Chengliang and Liu, Hongquan and Zhou, Shengjie and Chen, Hanghui},
  journal={Nat. Commun.},
  volume={16},
  number={1},
  pages={1054},
  year={2025},
  publisher={Nature Publishing Group UK London}
}

@article{PhysRevB.108.L140505,
  title = {Possible ${s}_{\ifmmode\pm\else\textpm\fi{}}$-wave superconductivity in ${\mathrm{La}}_{3}{\mathrm{Ni}}_{2}{\mathrm{O}}_{7}$},
  author = {Yang, Qing-Geng and Wang, Da and Wang, Qiang-Hua},
  journal = {Phys. Rev. B},
  volume = {108},
  issue = {14},
  pages = {L140505},
  numpages = {5},
  year = {2023},
  month = {Oct},
  publisher = {American Physical Society}
}

@article{PhysRevB.111.174506,
  title = {Effective model and pairing tendency in the bilayer Ni-based superconductor ${\mathrm{La}}_{3}{\mathrm{Ni}}_{2}{\mathrm{O}}_{7}$},
  author = {Gu, Yuhao and Le, Congcong and Yang, Zhesen and Wu, Xianxin and Hu, Jiangping},
  journal = {Phys. Rev. B},
  volume = {111},
  issue = {17},
  pages = {174506},
  numpages = {7},
  year = {2025},
  month = {May},
  publisher = {American Physical Society}
}

@article{10.1143/PTP.30.275,
    author = {Kanamori, Junjiro},
    title = "{Electron Correlation and Ferromagnetism of Transition Metals}",
    journal = {Progress of Theoretical Physics},
    volume = {30},
    number = {3},
    pages = {275-289},
    year = {1963}
}

@article{xia2025nickelate,
      title={Three-Dimensional Fermi Surface, Van Hove Singularity and Enhancement of Superconductivity in Infinite-Layer Nickelates}, 
      author={Chengliang Xia and Shengjie Zhou and Hanghui Chen},
      year={2025},
      journal={arxiv:2504.18778},
      archivePrefix={arXiv},
      primaryClass={cond-mat.supr-con}
}

@article{chen2026unified,
      title={A Unified Understanding of the Experimental Controlling of the T$_\text{c}$ of Bilayer Nickelates},
      author={Zeyu Chen and Jia-Heng Ji and Yu-Bo Liu and Ming Zhang and Fan Yang},
      year={2026},
      journal={arxiv:2603.14519},
      archivePrefix={arXiv},
      primaryClass={cond-mat.supr-con}
}

@article{zhang2025compr,
      title={Compressive Strain Turns $s^{\pm}$ into $d$-Wave Pairing in One-unit-cell La$_3$Ni$_2$O$_7$ Thin Film Via Substrate-Induced Hole Doping}, 
      author={Yang Zhang and Ling-Fang Lin and Adriana Moreo and Satoshi Okamoto and Thomas A. Maier and Elbio Dagotto},
      year={2025},
      journal={arxiv:2512.19520},
      archivePrefix={arXiv}
}

@article{Emery1995,
  title={Importance of phase fluctuations in superconductors with small superfluid density},
  author={Emery, V. J. and Kivelson, S. A.},
  journal={Nature},
  volume={374},
  number={6521},
  pages={434--437},
  year={1995},
  publisher={Nature Publishing Group UK London}
}

\end{document}